\documentclass[11pt,notitlepage]{revtex4-2}

\usepackage[utf8]{inputenc}
\usepackage{graphicx}
\usepackage{subcaption}
\usepackage{amsmath,amssymb}
\usepackage[normalem]{ulem}
\usepackage{xcolor}
\usepackage{hyperref}
\usepackage{booktabs}
\usepackage{float}
\graphicspath{{./figures/}}

\begin{document}

\title{An Overview of the MUSES Calculation Engine and How It Can Be Used to Describe Neutron Stars}

\author{Mateus Pelicer}
\email{mreinkep@kent.edu}
\author{Veronica Dexheimer}
\email{vdexheim@kent.edu}
\affiliation{Center for Nuclear Research, Department of Physics, Kent State University, Kent, OH 44240, USA}

\author{Joaquin Grefa}
\affiliation{Center for Nuclear Research, Department of Physics, Kent State University, Kent, OH 44240, USA}
\affiliation{Department of Physics, University of Houston, Houston, TX 77204, USA}


\begin{abstract}
For densities beyond nuclear saturation, there is still a large uncertainty in the equations of state (EoS) of dense matter that translate into uncertainties in the internal structure of neutron stars. The MUSES Calculation Engine provides a free and open-source composable workflow management system, which allows users to calculate the EoS of dense and hot matter that can be used, e.g. to describe neutron stars. For this work, we make use of two MUSES EoS modules, Crust Density Functional Theory and Chiral Mean Field model, with beta-equilibrium with leptons enforced in the Lepton module, then connected by the Synthesis module using different functions: hyperbolic tangent, Gaussian, bump, and smoothstep. We then calculate stellar structure using the QLIMR module and discuss how the different interpolating functions affect our results.
\end{abstract}

\maketitle


\section{Introduction}

Neutron stars are cold compact objects with layers that possess different compositions and cover different aspects of nuclear physics \cite{Baym:2017whm}. The outer crust, for example, contains neutron rich nuclei;  the inner crust is made of a liquid of deformed nuclei (including pasta shapes); and the core with bulk nuclear matter may contain exotic matter states such as hyperons, decuplet baryons, or deconfined quark matter. On the other hand, neutron stars only cover a very small part of the Quantum Chromodynamics (QCD) phase diagram being limited to the $T\approx 0$ axis. However, it is relevant source of information for dense cold nuclear matter, and astrophysical data can help in constraining the nuclear equation of state (EoS) \cite{MUSES:2023hyz}. 
At large temperatures, heavy-ion collisions are the main sources of experimental data to constraint the QCD phase diagram, in addition to lattice QCD simulations and perturbative QCD calculations, which serve as a theoretical benchmark for modeling and phenomenology \cite{MUSES:2023hyz}.

Given the multitude of different models that exist to describe strongly-interacting matter in different regions of the QCD phase diagram, the MUSES Calculation Engine (CE) was built in order to gather different descriptions (modules) of the nuclear EoS in a unique framework within MUSES \cite{MUSES_WEBSITE}. Each module is independent, but follows a standard format to make usage as simple as possible for users. 
MUSES stands for Modular Unified Solver of the Equation of State: it is modular because it allows different EoS descriptions that can be easily exchanged with one another since usage of all modules are standardized, and it is unified as the EoSs from different modules can be matched, resulting in a unified EoS that covers a broader range of the phase diagram than the original ones. There are also observable modules that can be used to connect EoSs with quantities that can be measured or observed. Furthermore, all MUSES software is free and open source.

On the neutron star side, currently, the available modules are: Crust-DFT (Density Functional Theory), Chiral Effective Field Theory ($\chi$EFT) and Chiral Mean Field (CMF) model modules. 
Crust-DFT is a phenomenological model that describes both nuclei, using DFT, and an interacting nucleon gas, using an excluded volume approach \cite{Du:2018vyp,Du:2021rhq,steiner_2025_14714273}. 
$\chi$EFT is the low-energy approximation of QCD, where nucleons are the degrees of freedom, and the EoS is computed via many-body perturbation theory \cite{Machleidt:2011zz,Drischler:2021kxf,zenodo_cheft}. Finally, CMF is a relativistic mean-field model based on chiral symmetry, that can describe the inner and outer core of the neutron stars and includes  the entire baryon octet, decuplet and/or quarks as degrees of freedom. In CMF, the baryons and quarks interact through the background meson fields, and a consistent deconfinement phase transition is implemented by a Polyakov-inspired field \cite{Dexheimer:2008ax,Dexheimer:2009hi,Cruz-Camacho:2024odu,zenodo_cmf}.

These EoS modules produce EoSs that are at least two-dimensional ($n_B$, $Y_Q$, where $Y_Q$ is the baryon and quark charge fraction) at $T=0$, for which the Lepton module can compute the $\beta$-equilibrated curve. These one-dimensional EoSs can then be matched together in the Synthesis module, responsible for combining the original EoSs within their overlapping regime of validity, either by a first-order phase transition (using a Maxwell or a Gibbs construction), or through an interpolation method that employs a sigmoid function.
The EoS modules can be used together with observable modules: Astrophysical observables can be computed using the QLIMR (meaning Quadrupole moment, Tidal Love number, Moment of Inertia, Mass, and Radius) \cite{Ravenhall1994,Bejger:2002ty,Yagi:2014qua,zenodo_qlimr}, which can be used to describe both static and slowly-rotating neutron stars; Out-of-equilibrium quantities such as bulk-viscosity and flavor relaxation time can be computed using the Flavor Equilibration module \cite{Alford:2023gxq,Alford:2024xfb,zenodo_flavor_equil}.

At the high $T$ side of the QCD phase diagram, the MUSES EoS modules relevant for heavy-ion collisions are constructed to reproduce lattice QCD results at vanishing $n_B$, the thermodynamics of the quark-gluon plasma at high $T$, and the crossover transition at low $\mu_B$. These models differ in how they incorporate phase transitions, critical phenomena, and hadronic degrees of freedom at low $T$. They also support varying dimensionalities and assumptions for chemical potentials and constraints such as net-strangeness neutrality or fixed electric charge fraction $Y_Q$, which are important for phenomenology. See Appendix \ref{HI} for more details. 

However, In this work, we defer a detailed exploration of the heavy-ion EoS modules to future work, and instead focus on the neutron star applications of the MUSES CE. In particular, we investigate how different matching schemes (hyperbolic tangent, Gaussian, bump, and smooth step) influence the resulting EoS and its impact on astrophysical observables, building on the previous analysis from Ref.~\cite{ReinkePelicer:2025vuh} using hyperbolic tangent only. We study how these merging or interpolation functions affect quantities such as the mass, moment of inertia, quadrupole moment, and tidal Love number of neutron stars.

\section{Formalism}

To connect different EoSs smoothly, one naturally resorts to sigmoid functions. For example, in ~\cite{Kambe:2016olv,Wang:2019npj},  the authors connect the hadron and quark EoSs through hyperbolic-tangent functions to simulate a crossover phase transition; in~\cite{Du:2018vyp}, the logistic function is used to connect EoSs valid in different density regimes; in~\cite{Albright:2015uua,PhysRevC.106.014909} a 'bump'-like function is used to match a Hadron Resonance Gas with a Quark-Gluon plasma EoS in the ($T, \mu_B$) plane. More recently, in 
a manuscript by the MUSES collaboration~\cite{ReinkePelicer:2025vuh}, the matching of 3 different EoSs was analyzed to build a unified neutron star EoSs using the hyperbolic-tangent in different thermodynamic variables. 

In this work we go beyond and analyze how different switching functions and different variables impact the EoS when matching a beta-equilibrated EoS with a crust (i.e. with nuclei) and another without, and how the matching function impacts neutron star observables. We use two of these Equations of State, Crust-DFT (I) and CMF (II). The switching is defined by some thermodynamic variable $Y$ as a function of another $x$ as
\begin{equation}\label{eq:interpolation}
Y(x) = Y^I(x) f_-(x) + Y^{II}(x) f_+ (x),
\end{equation}
where the variables $Y$ and $x$ can be chosen such that other thermodynamic quantities can be recovered. Here we focus on the cases $\varepsilon(n_B)$, $P(n_B)$, $P(\mu_B)$ and $c_s^2(n_B)$.  The interpolating functions are related by $f_+ (x) = 1 - f_-(x)$, and we consider four different cases for $f\pm$:
\begin{itemize}
    \item \textbf{Hyperbolic tangent}:
    \begin{equation}\label{eq:tanh}
    f_- (x) = \frac{1}{2} \left( 1 - \tanh{\left[ \frac{x - \bar x }{\Gamma} \right]} \right)\,,
    \end{equation}
    \item \textbf{Gaussian}:
    \begin{equation}
        f_-(x) =
    \begin{cases}
        1 & \text{if } x \leq \bar{x} \,,\\
        \exp\left[ - \left(\dfrac{x - \bar{x}}{\Gamma} \right)^n \right] & \text{if } x > \bar{x}\,,
    \end{cases}
    \end{equation}
    \item \textbf{Bump}:
    \begin{equation}
        f_-(x) =1 - \exp\left[- \left(\dfrac{\bar x}{x} \right)^n\right]\,,
    \end{equation}
    \item \textbf{Smoothstep:}
    \begin{equation}
        f_-(x) =
        \begin{cases}
            1 & \text{if } x \leq x_0 \\
            1 - 3t^2 + 2t^3 & \text{if } x_0 < x < x_1 \\
            0 & \text{if } x \geq x_1
        \end{cases}
        \quad\text{with } t = \dfrac{x - x_0}{x_1 - x_0}\,,
    \end{equation}
\end{itemize}
where $x_{0,1}$ are the endpoints of the interpolation region, and we relate them to $\bar x$ and $\Gamma$ as $x_0 = \bar x - \Gamma$ and $x_1 = \bar x + \Gamma$.

\begin{figure}[H]
    \centering
    \includegraphics[width=0.45\linewidth]{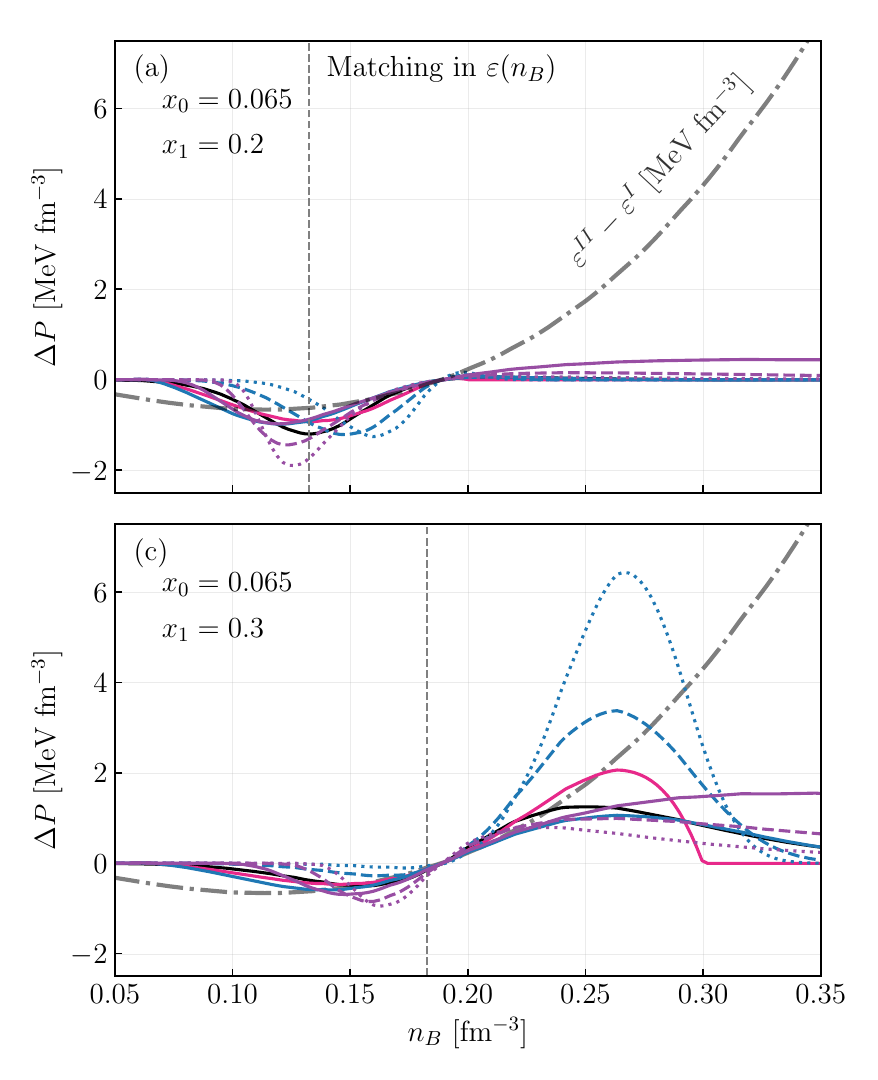}
    \includegraphics[width=0.45\linewidth]{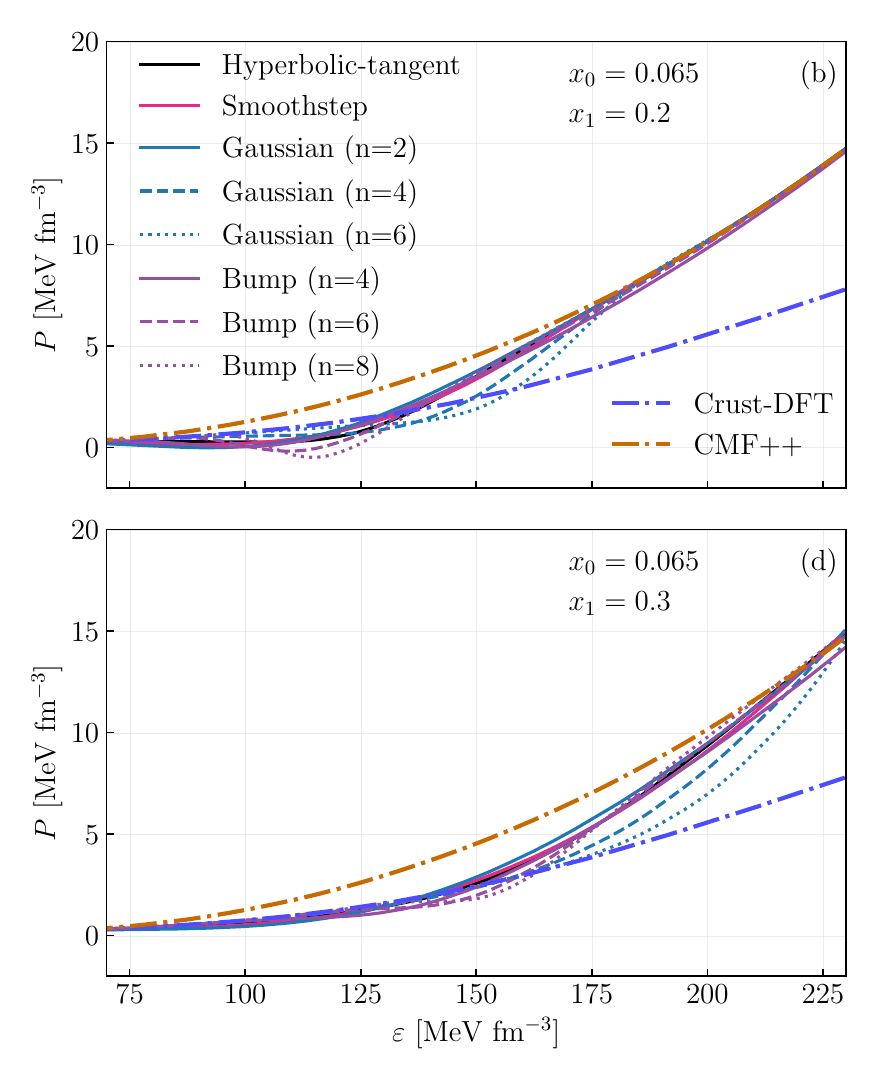}
    \caption{Rearrangement terms (panels a and c) and equation of state (panels b and d) for different interpolations with matching done in energy density as a function of density. The top and bottom panels have different endpoints for the interpolation region. The gray dotted-dashed curve indicates the difference in energy density between the two original equations of state, Crust-DFT and CMF.}
    \label{fig:e_interp}
\end{figure}

The parameters $\bar x$ and $\Gamma$ have different roles on the different functions: In the hyperbolic-tangent and in the smoothstep functions, $\bar{x}$ is the midpoint and $\Gamma$ is the width of the interpolation; in contrast, in the Gaussian, $\bar x$ marks the onset of the interpolating region (where $f_-$ starts to deviate from one), and $\Gamma$ controls the width.
For the bump function, $\bar x$ is the only shape-controlling parameter, which sets the both the endpoints and the width of the transition. The exponential power $n$ controls the sharpness of the transition in both the Gaussian and bump cases, with larger $n$ corresponding to narrower transitions. 
Both the Gaussian and bump-functions are assymetric with respect to the midpoint $x_{\rm mid} = (x_0+x_1)/2$. The Gaussian is broader and left-skewed for n=2, but it becomes more right-skewed at higher $n$ due to the transition growing steeper closer to $x_1$. The bump function is steep around the midpoint, but has a larger tail, which is more persistent at lower $n$. 
Note that the smoothstep is a $C^1$ function: it is continuous in the first derivative, but introduces discontinuities in the second. Consequently, it induces a second-order phase transition at the boundaries. While this is sufficient for obtaining basic thermodynamic variables, it is not a recommended method if the analysis of susceptibilities is of interest. However, one can always devise functions that are differentiable to higher order, or use the other functions discussed here if high-order susceptibilities are of interest.

\begin{figure}[t!]
    \centering
    \includegraphics[width=0.45\linewidth]{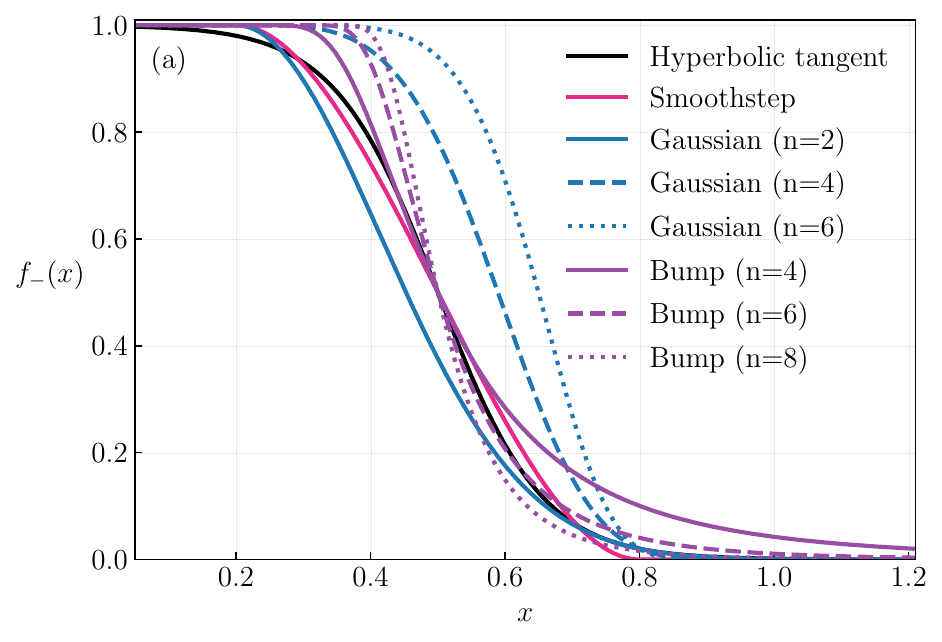}
        \includegraphics[width=0.45\linewidth]{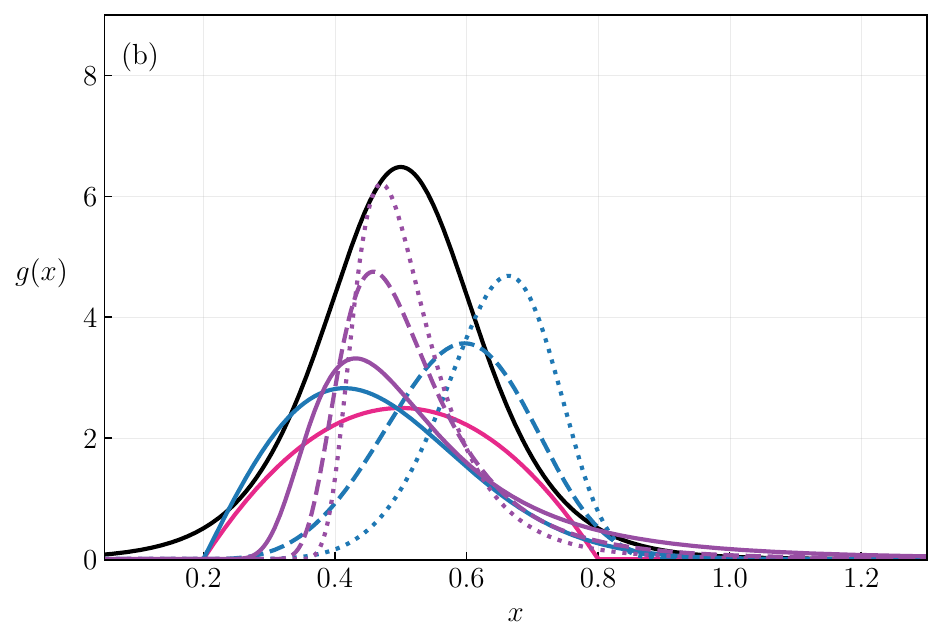}
    \caption{Interpolating functions  (panel a) and derivatives (panel b) for the 4 cases considered in this work, with $n$ controlling the sharpness of the transition.}
    \label{fig:f_g}
\end{figure}

The features discussed above are illustrated in Figure \ref{fig:f_g}, panel a), which shows $f_-$ for each interpolation function. For the Gaussian function, we show $n=2, 4, 6$, and for the bump function we show $n=4, 6, 8$, where  $n=2$ makes the interpolation region too broad. Due to the different role of ($\bar x,\; \Gamma$) in the different functions, comparing them with a common  set of values is not meaningful. Instead, we choose the parameters by defining two points for the interpolation region, $x_0$ and $x_1$, that are related to its beginning and the end, respectively. For the hyperbolic-tangent function, we set both parameters from the conditions $f_-(x_0) = 1-\delta$ and $f_-(x_1) = \delta$. For the Gaussian function, we set $\bar x = x_0$, to match the starting of the interpolating region and $\Gamma$ to satisfy $f_-(x_1) = \delta$. 
For the bump function, we choose $\bar x$ to set the midpoint of the transition via $f_-(x_{\rm mid}) = 1/2$. Notice though that the bump function, like the Gaussian, is asymmetric w.r.t. $x_{\rm mid}$.
The smoothstep is constrained by the endpoints by definition.
Note that a smaller value of $\delta$ results in a smaller $\Gamma$ for fixed endpoints, as the function must be squeezed in the same interval to meet the defining criteria. We choose $\delta=0.02$, which we found, by trial and error: larger values allow the transition to spread too much outside the endpoints, and smaller values are too restrictive and produce unstable EoSs. For the endpoints, we take  $x_0=0.2$ and $x_1=0.8$, in Figure~\ref{fig:f_g}.

\begin{table}[t!]
\caption{Location and value of the maximum of the derivative $g(x)$ for different interpolation functions. The smoothstep function is defined over a compact interval, with its maximum also at the midpoint.}
\renewcommand{\arraystretch}{1.5}
\centering
\begin{tabular}{lcc}
\toprule
\textbf{Function} & \textbf{Location of $g_{\rm max}$} & \textbf{Value of $g_{\rm max}$} \\
\midrule
Hyperbolic tangent & $x = \bar{x}$ & $g_{\rm max} = 1/2\Gamma$ \\
Gaussian & $x = \bar{x} + \Gamma \left( \dfrac{n-1}{n} \right)^{1/n}$ & 
$g_{\rm max} = \dfrac{n}{\Gamma} \left( \dfrac{n-1}{n} \right)^{\frac{n-1}{n}} e^{-(n-1)/n}$ \\
Bump & $x = \bar{x} \left( \dfrac{n}{n+1} \right)^{1/n}$ & 
$g_{\rm max} = \dfrac{n}{\Gamma} \left( \dfrac{n+1}{n} \right)^{\frac{n+1}{n}} e^{-(n+1)/n}$ \\
Smoothstep & $x = \bar{x}$ & $g_{\rm max} = 3/2\Gamma$ \\
\bottomrule
\end{tabular}
\label{tab:g}
\end{table}

The dependence of the interpolating functions $f_\pm$ on $x$ gives rise to rearrangement terms (i.e. corrections to the variables computed from thermodynamic relations), which depend on the derivative 
\begin{equation}
    g(x) = \pm \frac{d f_\pm}{dx}.
\end{equation}
The integral of $g(x)$ over the entire real axis is one. Therefore, $g(x)$ presents a large peak value for transitions that occur in a narrower region, and smaller peaks for broader transitions. This leads to large rearrangement terms in the interpolated quantities. Additionally, the rearrangement is enhanced if the EoSs being interpolated differ significantly. See Eqs.~(40)--(52) of~\cite{ReinkePelicer:2025vuh} for explicit expressions of the rearrangement terms, which are the same for any choice of $f_\pm(x)$.

In panel b) of Figure \ref{fig:f_g} we show $g(x)$, using the same parameters as in panel a).  The hyperbolic-tangent has the largest peak for $g(x)$, which corresponds to it being the steepest transition, while the smoothstep has a smaller maximum value but is larger closer to the endpoints. The bump function has the distinct feature of having a very long tail for the rearrangement. For n=4, particularly, the rearrangement reaches $~10^{-6}$ only for $x \approx 20 \bar x$. See Table \ref{tab:g} for the maximum value of $g(x)$ and its location. 

Figure \ref{fig:e_interp} shows the rearrangement term $\Delta P = g(n_B) n_B(\varepsilon^{II} - \varepsilon^{I})$ for the $\varepsilon(n_B)$ interpolation for $x_0 = 0.065$ and $x_1=0.2$ (panel a) and $x_1=0.3$ (panel c) that arises in the $\varepsilon(n_B)$ interpolation. Because the energy density in the Crust-DFT model is larger than in CMF in the region of $n_B\lesssim 0.18$ fm$^{-3}$, the pressure correction is negative. Because the pressure is already low at such densities, the interpolation can lead to negative pressures and thus to unstable phases. This is confirmed by the corresponding EoSs, shown in panel b), which shows that most interpolating functions lead to unstable phases, except the Gaussian with n=4 and n=6. This is because they are right-skewed w.r.t. to the midpoint, and thus pick up a correction smaller in magnitude.  In panels c) and d), we show the same quantities as a) and b), respectively, but we allow the transition in a broader region (from $n_B=0.065$ to $0.3$ fm$^{-3}$). The broadening reduces the peak value of the rearrangement, and reduces the magnitude of the correction at low densities, no longer leading to negative pressures. However, the higher densities covered lead to a positive pressure correction, as the energy difference between EoSs changes sign, and the steep increase leads to higher rearrangement terms, despite the smaller peak in $g(x)$.

\begin{figure}[t!]
    \centering
    \includegraphics[width=0.45\linewidth]{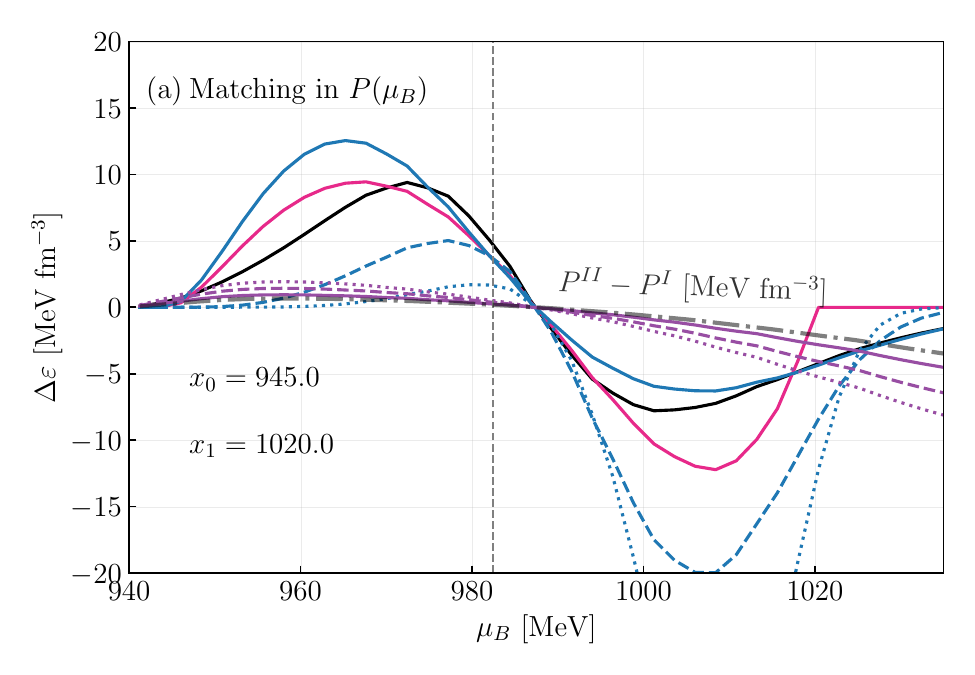}
    \includegraphics[width=0.45\linewidth]{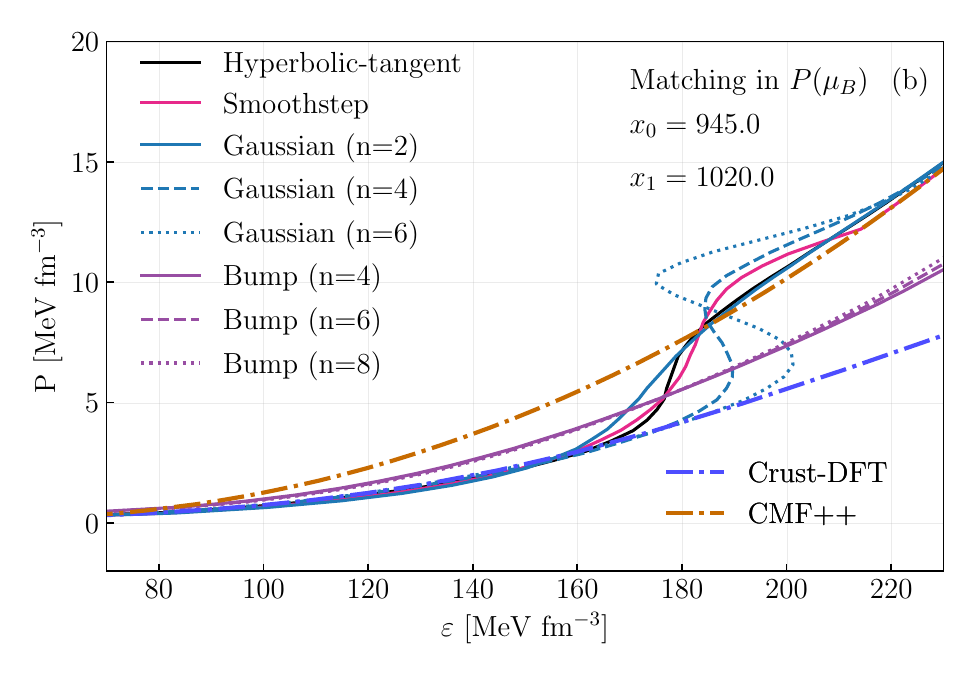}
    \caption{Rearrangement terms (panel a) and equation of state (panel b) for different interpolations with matching done in pressure as a function of baryon chemical potential. The gray dotted-dashed curve indicates the difference in pressure between the two original equations of state, Crust-DFT and CMF.}
    \label{fig:pmub_interp}
\end{figure}

In Figure~\ref{fig:pmub_interp} we show the rearrangement $\Delta \varepsilon$ (left) and the EoS (right) for the $P(\mu_B)$ interpolation. The rearrangement is given by $\Delta \varepsilon = g(\mu_B) \mu_B \left(P^{II} - P^I \right)$. In this case, the difference between the pressure of the models (gray curve) is not significantly large -- less than 5 MeV fm$^{-3}$ in magnitude -- however, because the rearrangement is multiplied by the chemical potential, the total correction becomes significant. The role of the rearrangement is to make the region where $\Delta \varepsilon >0$ softer, as the corresponding interpolated pressure corresponds to a larger $\varepsilon$, while for $\Delta \varepsilon<0$, the EoS becomes stiffer, as the interpolated pressure corresponds to a smaller $\varepsilon$. This is exemplified by the region with $\varepsilon$ between 180 and 200 MeV fm$^{-3}$, where the interpolated EoS has a larger pressure than both EoSs I and II. Because the Gaussians with n=4 and 6 are right-skewed, they lead to an unstable EoS, due to the matching being concentrated in a smaller interval than the other methods, and also to a region where the pressure difference is larger. This is the same as discussed in Figure~\ref{fig:e_interp}. 

While the behavior of the rearrangement follows the same overall pattern for all functions, there are some caveats: the hyperbolic-tangent, smoothstep and Gaussian (n=2) are more contained within the interpolation region, with the Gaussian being the largerst at low densities and the smallest at high densities. The Gaussians with n=4, 6 and the bump functions have a small rearrangement close to starting point, but grow with increasing density (or chemical potential), with the Gaussians having a large magnitude close to the ending point, and the bump functions growing beyong the endpoint, due to the very broad tail of $g(x)$. The persistence of the rearrangement for the bump functions is more pronounced in the $P(\mu_B)$ plane due to existence of a single parameter $\Gamma$ that controls its behavior. At large $x$, $g(x)$ decays as a power law, and having an absolute value of $\bar x \sim 1000$ leads to a very broad rearrangement, in comparison to $x \sim 0.1$, justifying the different behavior in the $\varepsilon(n_B)$ and $P(\mu_B)$ interpolations.

\begin{figure}[t!]
    \centering
    \includegraphics[width=0.45\linewidth]{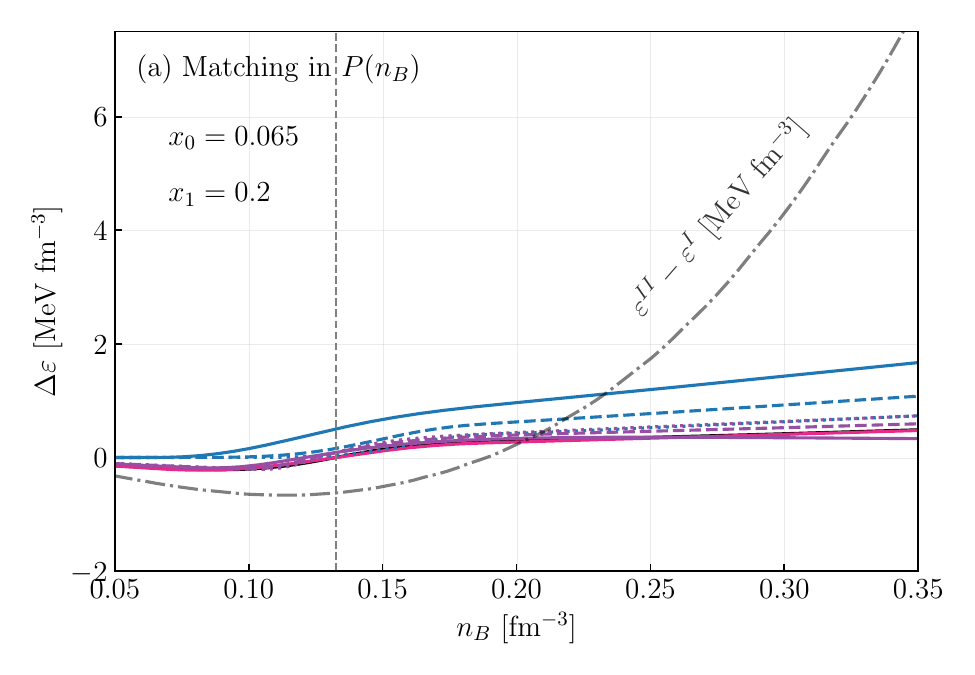}
    \includegraphics[width=0.45\linewidth]{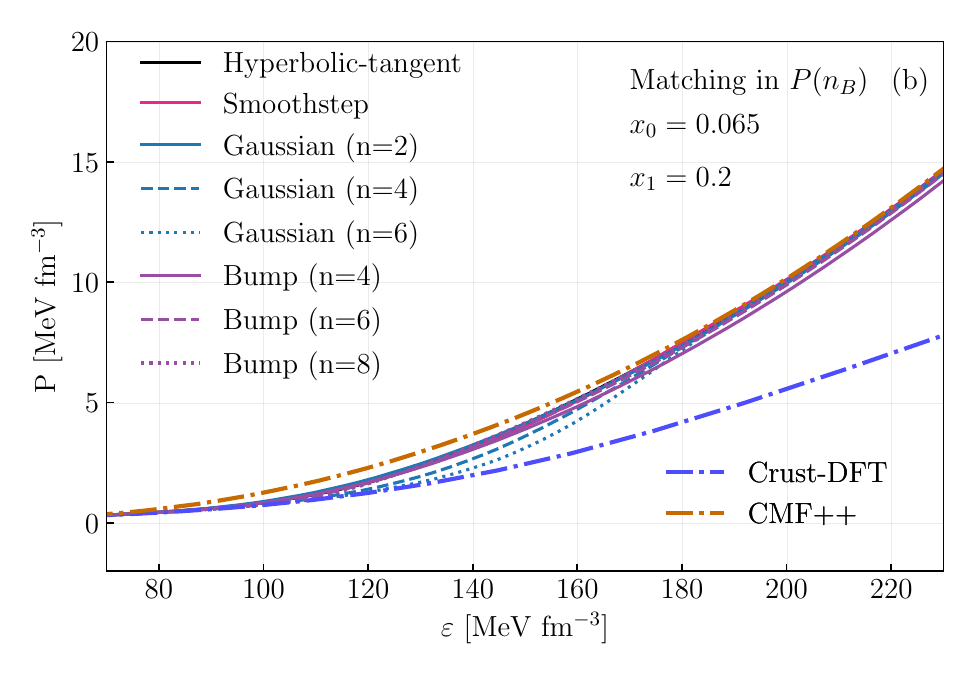}
    \caption{Rearrangement terms (panel a) and equation of state (panel b) for different interpolations with matching done in pressure as a function of density. The gray dotted-dashed curve indicates the difference in energy density between the two original equations of state, Crust-DFT and CMF.}
    \label{fig:pnb_interp}
\end{figure}

If instead we choose the interpolation in the $P(n_B)$ plane, the rearrangement term becomes an integral, and can be considered a global correction
\begin{equation}
\Delta \varepsilon = n_B \int^{\bar n_B}_{n_B} d n_B' g(n_B')\left(\varepsilon^{II} - \varepsilon^{I}\right)/n_B'\,,
\end{equation}
as it is not restricted to the matching region, being finite even for $n_B \gg \bar x + \Gamma$. This is shown in panel a) of Figure~\ref{fig:pnb_interp}. In this matching case, the rearrangement is small compared to the $\varepsilon(n_B)$ and $P(\mu_B)$, as the integration is less dependent on rapid changes in the functions $g(x)$ and $\varepsilon^{II} - \varepsilon{I}$. As a drawback, at high $n_B$ we only recover EoS II in the $P(n_B)$ plane, but because $\varepsilon$ is shifted by the rearrangement term, then it becomes slightly shifted in the $P(\varepsilon)$ plane w.r.t. to the original one, as exemplified in panel b).

\begin{figure}[t!]
    \centering
    \includegraphics[width=1\linewidth]{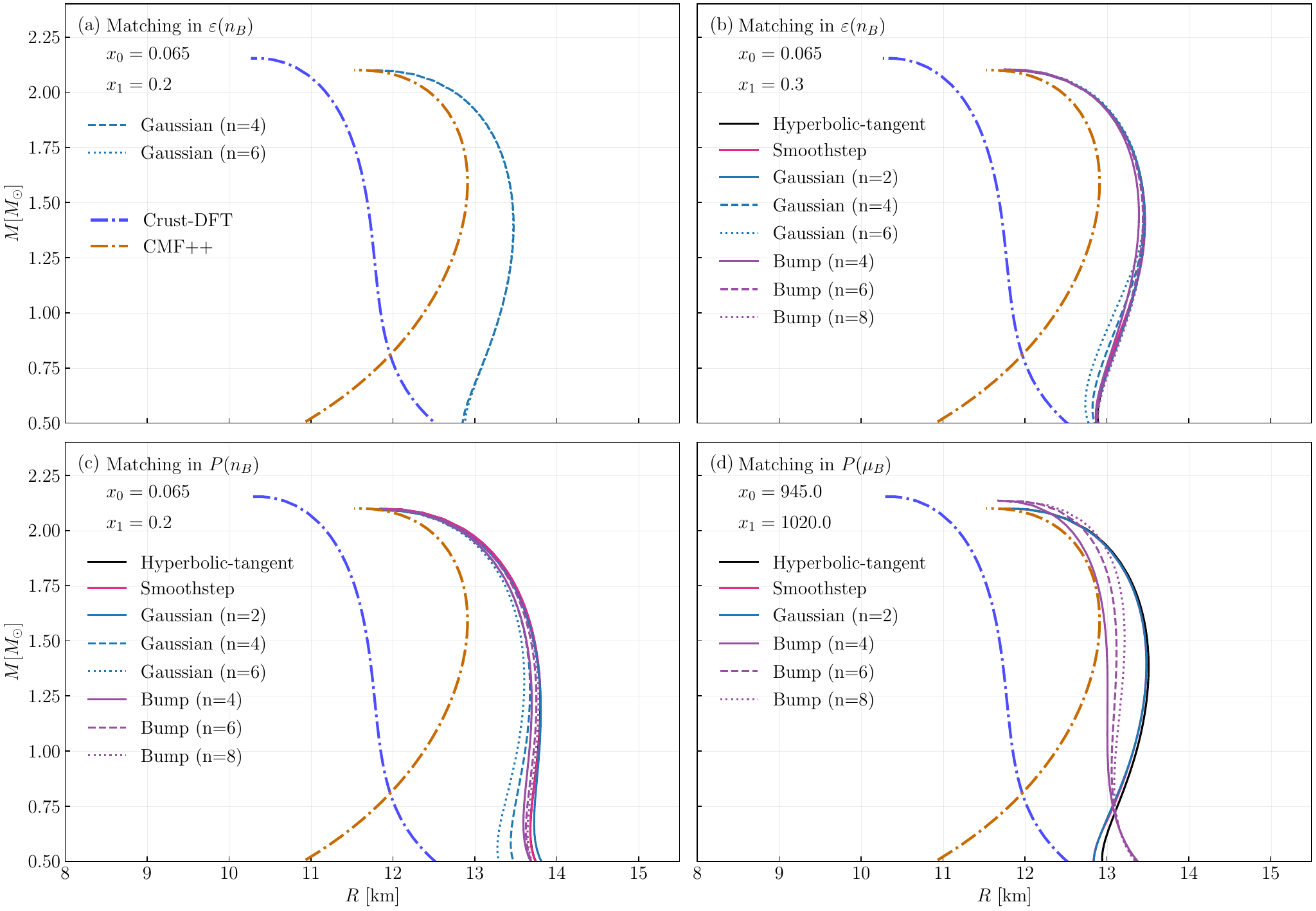}
    \caption{Mass-radius diagram for different interpolated equations of state that are stable, along with the original crust-DFT and CMF ones. Each panel shows the  matching done a different thermodynamic variable: energy density as a function of density (panels a and b, for different endpoints), pressure as a function of density (panel c), and pressure as a function of baryon chemical potential (panel d).}
    \label{fig:m_r}
\end{figure}

In Figure~\ref{fig:m_r}, we show the mass-radius curves computed by solving the TOV equation for static stars in the MUSES module QLIMR with the interpolated EoSs that are thermodynamically stable, alongside the Crust-DFT and CMF (without crust) models. All the matched EoSs approximately preserve the maximum mass predicted by the CMF model, with the exception of the bump functions in the $P(\mu_B)$ interpolation. In that case, the long tail of the rearrangement modifies the high-density region of the interpolated EoS, leading to an increase of the maximum mass. 

The radius of the canonical 1.4$M_\odot$ stars only show modest variations in the interpolated EoSs, ranging between 13 and 14~km. If we exclude the bump functions in the $P(\mu_B)$ case, the variation is even smaller, with $13.5<R<13.8$~km. 
Comparing the radii of the interpolated EoSs with the original CMF curve we see an increase in radius of $\sim$~500--800m due to the inclusion of the crust. Additionally, varying the endpoints of the interpolation in the $\varepsilon(n_B)$ plane between 0.2~fm$^{-3}$ and 0.3~$fm^{-3}$ produces very similar curves, despite the different matching functions. This indicates that if the matching occurs above the crust-core transition, the influence of the specific matching point is subdominant relative to the overall EoSs behavior. This suggests that, if the matching occurs around the crust-core transition, the precise location of the matching point plays a subdominant role in determining the properties of the star. While the matching procedure does introduce some systematic uncertainty, its overall impact on macroscopic observables, such as the radius of a 1.4~$M_\odot$ star and the maximum NS Mass, is relatively small.

\begin{figure}[t!]
    \centering
    \includegraphics[width=0.9\linewidth]{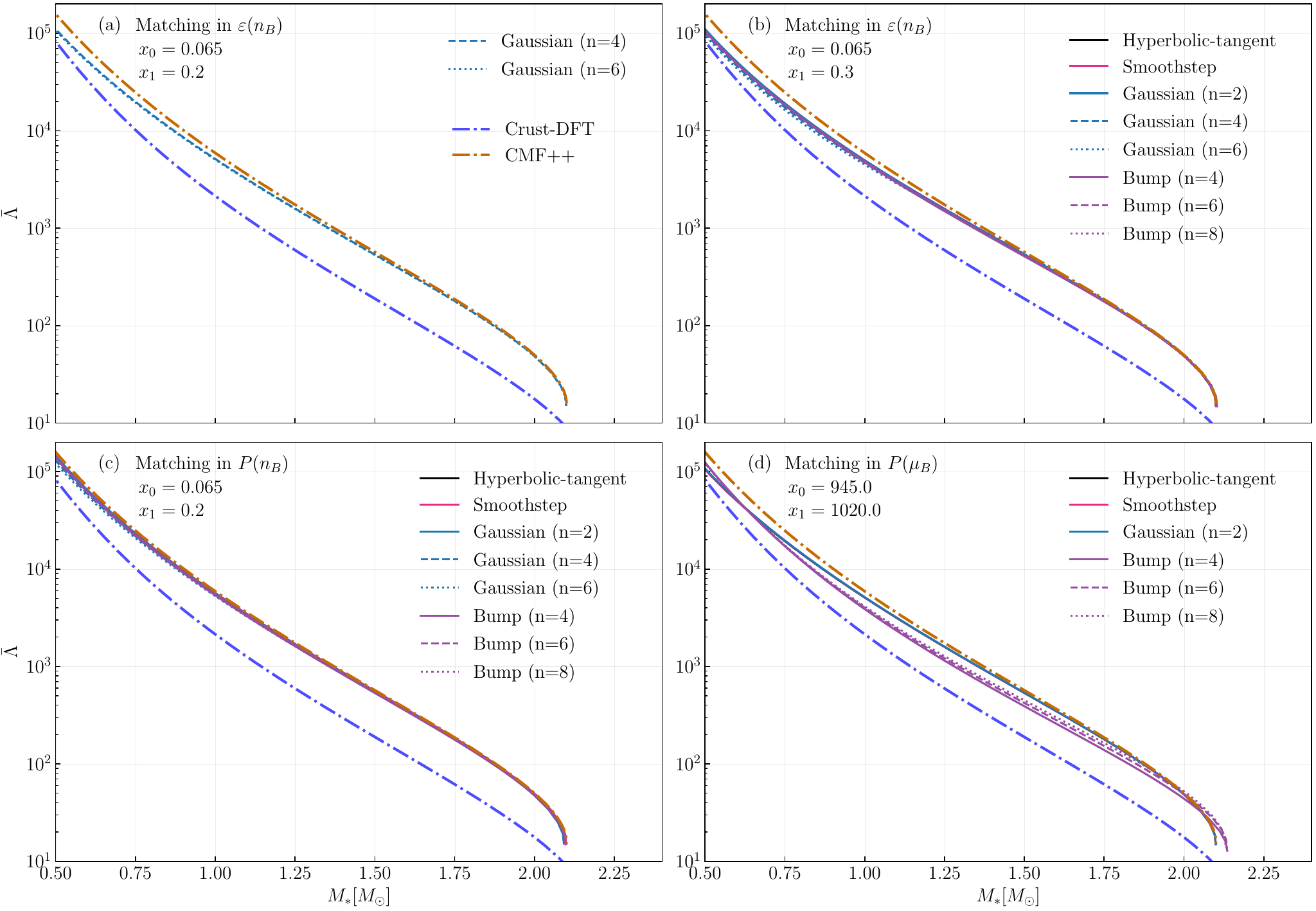}
    \caption{Dimensionless tidal deformability-radius diagram for different interpolated equations of state that are stable, along with the original crust-DFT and CMF ones. Each panel shows the  matching done a different thermodynamic variable: energy density as a function of density (panels a and b, for different endpoints), pressure as a function of density (panel c), and pressure as a function of baryon chemical potential (panel d).}
    \label{fig:l_m}
\end{figure}

Also from the QLIMR module, in Figure~\ref{fig:l_m}, we show the dimensionless tidal deformability $\bar \Lambda = \Lambda/M^5$. As with the M-R curves, different endpoints in the $\varepsilon(n_B)$ matching predict very similar results: both sets of curves match the CMF predictions for stars with $M>1.5M_\odot$, and  interpolate smoothly down to the Crust-DFT curve at lower mass. The matching in the $P(n_B)$ plane maintains agreement with the CMF prediction down to $~0.5M_\odot$, possibly due to the small rearrangements contributions to the EoS in this case. Finally, the $P(\mu_B)$ case are nearly identical to the $\varepsilon(n_B)$ curves, except for the bump curves, which stay in-between the Crust-DFT and CMF.

\section{Discussion and conclusions}

In this work, we use the MUSES cyberinfrastructure to produce beta equilibrated neutron stars combining the Crust Density Functional Theory (crust-EFT) and Chiral Mean Field (CMF) model to describe different density regimes, the former for the lower density and the latter for the higher density regime. We matched the two equations of state using different interpolating schemes, hyperbolic tangent, Gaussian, bump, and smoothstep.
In each case, we vary the interpolation parameters to study their effects on the resulting equation of state (EoS). We also test, for each interpolation scheme, performing the interpolation over different thermodynamic variables. We find that the transitions that were made over energy density as a function of density, $\varepsilon(n_B)$, and pressure as a function of baryon chemical potential, $P(\mu_B)$, over a narrow interval result in unstable EoSs due to negative rearrangement terms. This is not an issue for transitions over $P(n_B)$, although in this case the original high density EoS is not recovered in all thermodynamic variables.

Overall, the tidal deformability and the mass-radius are consistent across most interpolating functions, and variations are within current observational uncertainties~\cite{MUSES:2023hyz}. These results suggest that, provided the matching is performed between the crust-core transition, the specific interpolation scheme introduces only modest uncertainty in macroscopic observables. Therefore, crust-core matching does not compromise the ability to compare EoSs to astrophysical constraints from gravitational-wave observations or mass-radius measurements. 
From the small variations obtained, the most noticeable ones appeared in interpolations performed with a bump function over $P(\mu_B)$. This is not surprising, as the rearrangement term modifies the EoS up to densities reaching the center of the star.
Additionally, these rearrangements terms introduce mild bumps in the speed of sound, which have been shown to affect neutron star properties, including mass, radius, and tidal deformability \cite{Tan:2020ics,Tan:2021ahl,Tan:2021nat}. However, the EoSs generated with the interpolating functions are not as extreme as those discussed in the references, leading to moderate variation in the observables.


\appendix
\section{Heavy-ion Collisions}
\label{HI}

In heavy-ion simulations, the equation of state (EoS) serves as the core input for hydrodynamic evolution. It must account for thermal equilibrium while allowing for local fluctuations in conserved charges (baryon number $B$, strangeness $S$, and electric charge $Q$) \cite{ALICE:2016fzo,Plumberg:2024leb}. A realistic EoS should therefore be defined over a 4D space of inputs: temperature $T$, and chemical potentials $\mu_B$, $\mu_Q$, and $\mu_S$. Key output variables include pressure $P$, entropy density $s$, energy density $\varepsilon$, baryon density $n_B$, and the speed of sound squared $c_s^2$. The finite $T$ MUSES modules also provide susceptibilities, relevant for fluctuations of conserved charges and the study/search for the QCD critical point.

The three heavy-ion EoS modules currently supported in MUSES are: 4D Taylor-expanded Lattice ($BQS$), Ising 2D $T'$-Expansion Scheme (TExS), and the Holographic (NumRelHolo). 
$BQS$, based on lattice QCD data, is a module that provides an EoS obtained from a Taylor series expansion in $\mu_B/T$, $\mu_Q/T$, and $\mu_S/T$, valid at high $T$ and low chemical potentials. It includes hadronic contributions via the Hadron Resonance Gas (HRG) at low $T$ and only exhibits a smooth crossover without a critical point~\cite{Noronha-Hostler:2019ayj, jahan_2025_14639786}. 
As a 2D model in $T$ and $\mu_B$, TExS matches lattice QCD based on the novel $T'$~-expansion scheme \cite{Borsanyi:2021sxv} and incorporates a 3D Ising model universality class to include a critical point and a first-order phase transition at large $\mu_B$. It can be run with or without a critical point ~\cite{Kahangirwe:2024cny,kahangirwe_2025_14637802}. 
Inspired by the gauge/gravity duality, NumRelHolo is a bottom-up holographic model that captures strongly coupled QCD-like thermodynamics. It is constrained to reproduce the lattice results at vanishing density, describes a crossover at low $\mu_B$, agrees with the state-of-the-art lattice thermodynamics at finite density, and predicts the location of the QCD critical point and the first-order phase transition at higher $\mu_B$ ~\cite{Critelli:2017oub,Rougemont:2023gfz,Grefa:2021qvt,Hippert:2023bel,yang_2025_14695243,hippert_2024_13830379}.
Additionally, transport coefficients—crucial for understanding non-equilibrium dynamics—are under development or available in select models. 


\section*{Funding}
This work was supported in part by the National Science Foundation (NSF) within the framework of the MUSES collaboration, under grant number OAC-2103680.
This work used Jetstream2 at Indiana University and Open Storage Network at NCSA through allocation PHY230156 from the Advanced Cyberinfrastructure Coordination Ecosystem: Services \& Support (ACCESS) program \citep{NSFACCESS}, which is supported by National Science Foundation grants \#2138259, \#2138286, \#2138307, \#2137603, and \#2138296.

\section*{Data Availability}
The data from \cite{ReinkePelicer:2025vuh} is already reproducible from the MUSES CE \cite{muses_calculation_engine_2025_v1}. All data shown in this work will be reproducible using the MUSES cyberinfrastructure \cite{muses_calculation_engine_2025_v1} soon.

\section*{Acknowledgments}
We would like to thank the wider MUSES collaboration for many discussions during our collaboration meetings and all colleagues who helped with testing the MUSES CE.



\section*{Abbreviations}
The following abbreviations are used in this manuscript:\\

\noindent 
\begin{tabular}{@{}ll}
QCD & Quantum Chromodynamics\\
EoS & Equation of State\\
CE & Calculation Engine \\
MUSES & Modular Unified Solver of the Equation of State \\
CMF & Chiral Mean Field \\
DFT & Density Functional Theory
\end{tabular}

\bibliography{inspire}

\end{document}